# Enabling rapid and accurate construction of CCSD(T)-level potential energy surface of large molecules using molecular tailoring approach

Subodh S. Khire,[1] Nalini D. Gurav,[1] Apurba Nandi,[2*] Shridhar R. Gadre[1*]

[1]Department of Scientific Computing, Modelling and Simulation, Savitribai Phule Pune University, Pune 411 007, India.

[2]Department of Chemistry and Cherry L. Emerson Center for Scientific Computation, Emory University, Atlanta, Georgia 30322, U.S.A.

gadre@unipune.ac.in, gadre@iitk.ac.in

**Abstract**

The construction of the potential energy surface (PES) of even a medium-sized molecule employing correlated theory, such as CCSD(T), is an arduous task due to the high computational cost. In this Letter, we report the possibility of efficient construction of such a PES employing the molecular tailoring approach (MTA) on off-the-shelf hardware. The full calculation (FC) as well as MTA energies at CCSD(T)/aug-cc-pVTZ level for three test molecules, viz. acetylacetone, N-methyacetamideand and tropolone are reported. All the MTA energies are in excellent agreement with their FC counterparts (typical error being sub-millihartree) with a time advantage factor of 3 to 5. The energy barrier from the ground- to transition-state is accurately captured. Further, the accuracy and efficiency of the MTA method for estimating energy gradients at CCSD(T) level are demonstrated. This work brings out the possibility of the construction of PES for large molecules using MTA with the computational economy at a high level of theory and/or basis set.

**TOC Graphics**

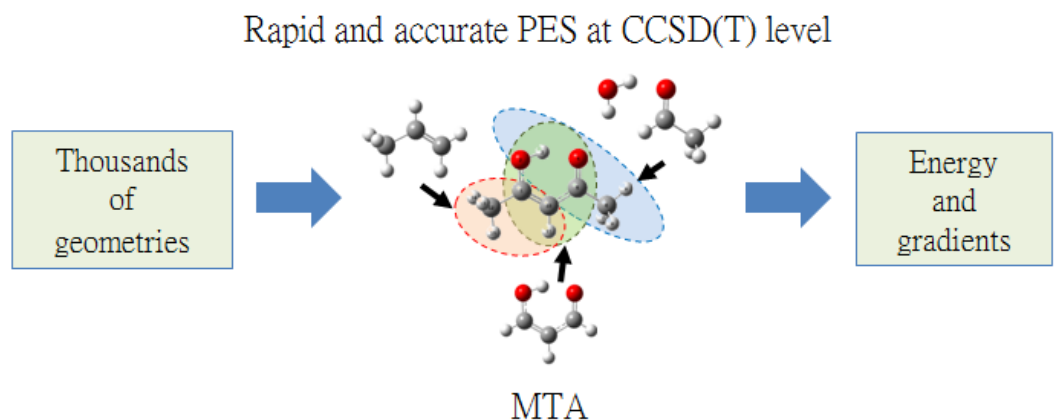

**Introduction:**

Developing high-dimensional potential energy surfaces (PES) based on advanced level *ab initio* methods continues to be an active area of theoretical and computational chemistry. Within the Born-Oppenheimer approximation, one can get the PES "on the fly" by obtaining the electronic energies from electronic structure packages. However, the computational cost quickly becomes prohibitive as the level of electronic structure theory as well as the system size goes up. An alternative approach is to develop an accurate analytical representation of PES using data sets of electronic energies that span a high dimensional space. As the PES establishes a relationship between a nuclear configuration and the forces acting on nuclei, it has wide applications in chemical/reaction dynamics, molecular spectroscopy etc. The PES thus constructed have been employed in the literature for several types of reactions dynamics calculations, harmonic-anharmonic vibrational analysis, geometry optimization etc.

In the past 15 years, significant progress has been made in the development of nonparametric, machine-learning approaches for fitting large data sets of electronic energies for molecules and clusters that contain more than four atoms. Three major methods recently in widespread use for this purpose are: permutationally invariant polynomials (PIPs), Neural Networks (NN), and Gaussian Process Regression (GPR) as well as their combinations. A number of reviews have appeared that summarize these approaches.[1–9] However, it is a major challenge to extend these methods to large molecules of interest in chemistry and material science.

Though accurate, the quantum chemical methods capturing a substantial part of electron correlation such as second-order Møller-Plesset second order perturbation theory (MP2), coupled-cluster with singles and doubles (CCSD), coupled-cluster with perturbative triples [CCSD(T)] etc. have a formidable scaling, leading to the requirement of high processor speed, memory, and secondary storage. Thus, there is a bottleneck for developing the PES at high-level theory as the molecular size increases. Due to the very high scaling of the "gold standard" CCSD(T) theory ($\sim N^7$, $N$ being the number of basis functions), making it unsuitable for systems containing more than 10 atoms. A large data set of energies is required for high dimensional PES and this data set requirement rapidly increases with the increasing of number of atoms. Therefore, researchers have been focusing on simultaneous energy plus gradient fitting method for the substantial reduction of the fitting data points. A major advantage of using energy plus gradient data is found for small configuration data sets where the PES exhibits equivalent fidelity to that obtained using only energy data sets, which are an order of magnitude larger. For dealing with PES of larger molecules, lower-level electronic structure methods such as density functional theory (DFT) and MP2 are normally used. The PES of molecules having more than 10 atoms using CCSD(T) level of theory are generally conspicuous by their absence. In 2016, Bowman and co-workers computed the PES for the formic acid dimer $(HCOOH)_2$, a 10-atom system, using 13475 energies at CCSD(T)-F12/haTZ (VTZ for H and aVTZ for C and O) level of theory.[10] This PES was applied for zero-point energy (ZPE) computation using Diffusion Monte-Carlo (DMC) method and ground-state tunneling splitting for H-transfer process. A 9-atom PES for the chemical reaction $Cl+C_2H_6$ was recently reported by Papp *et al.* using a composite MP2/CCSD(T) method.[11] Examples of potentials for 6 and 7-atom chemical reactions which are fits to tens of thousands or even hundred thousand CCSD(T) energies have also been reported.[12-13] Recently, a 15-atom PES was calculated for Acetylacetone, which contains 7 heavy atoms.[14] This was a major computational effort at the LCCSD(T)-F12/aug-cc-pVTZ level of theory. This PES was obtained with 2151 LCCSD(T) energies using Δ-Machine Learning approach.

As pointed out earlier, *ab initio* methods such as DFT and MP2, do not yield accurate results for reaction barriers, geometrical parameters and vibrational frequencies compared to the their CCSD- or higher counterparts. Thus, there is a need to generate a PES based on CCSD- or CCSD(T) level calculations. Nevertheless, developing the PES using CCSD(T) level of theory for molecules containing more than 10 atoms is prohibitively expensive.

In order to circumvent this high scaling problem, several methods have been proposed during the last two decades for treating large molecules by fragmentation-based (FB) approach. Some of these methods have been extensively tested and benchmarked for obtaining the molecular energies, gradients and Hessian matrix. An overview of FB methods, enabling *ab initio* calculations on large molecules/clusters methods has been given in Ref. 15 and 16. In this Letter, we present a first-ever, well-benchmarked application of an FB method, viz., molecular tailoring approach (MTA) for generating the PES of large molecules of interest at CCSD(T) level theory. The method MTA was proposed, developed and tested out on a variety of molecules/clusters by Gadre and co-workers.[17-22] The MTA framework is currently applicable to closed-shell systems for estimating the energies/gradients/Hessian matrix. Geometry optimization followed by IR/Raman spectral calculation is also available within the MTA. We briefly capture the essential features of MTA below.

Within MTA, a spatially extended, closed-shell parent molecule under investigation is divided into a series of primary and overlapping fragments. These fragments are subjected to computation instead of the entire parent molecule. The MTA program 'patches' the results (e.g. Energy, E) of the fragments to get an excellent approximation to the property of the parent molecule. Eq. (1) employing set inclusion/exclusion principle.

$$E = \sum_i E^{F_i} - \sum_{i<j} E^{F_i \cap F_j} + \cdots + (-1)^k \sum_{i<j<\cdots<n} E^{F_i \cap F_j \cap \cdots \cap F_n} + \cdots \qquad \ldots(1)$$

Here, $E^{F_i}$ represents the energy of i[th] primary fragment while $E^{F_i \cap F_j}$ is the energy of overlap fragment between i[th] and j[th] primary fragments and so on, $k$ being the degree of overlap. The term $n$ represents the number of primary fragments.

On account of this, the required computational expenses/resources for estimating a property of parent molecule, are steeply reduced. Within the MTA procedure, initial fragments are made by putting a sphere on all non-hydrogen atoms. These fragments are merged based on distance, with subsequent rounds of fragmentation. Thus, the near-neighbourhood of an atom is preserved in at least one of the main fragments. The quality of the results can be gauged a priory by the distance-based parameter, called as R-Goodness (RG). In general, larger is the RG, the better is the preservation of the chemical environment of atoms. For more details about MTA's origin, fragmentation procedure and capabilities, and limitations of MTA, readers are directed to Ref. 17-21.

In 2012, a grafting procedure was devised and thoroughly benchmarked [19,20] in order to reduce the error arising due to the fragmentation. Within the version of grafting procedure used here for calculating the energy,[19] we take the difference between the Full Calculation (FC) and MTA correlation energies computed with lower basis (LB). This difference is added to the sum of MTA correlation energy at higher basis (HB) and the Hartree-Fock (HF) FC energy at HB. All the MTA calculations are done using identical fragmentation scheme, *vide* Eq. (2). In the following discussion, we call such grafted energy (*cf.* Eq. (2)) as the GMTA energy.

$$E_{GMTA} = E(HF)_{FC}^{HB} + E(CORR)_{MTA}^{HB} + (E(CORR)_{FC}^{LB} - E(CORR)_{MTA}^{LB}) \qquad \ldots(2)$$

Here, $E(HF)_{FC}^{HB}$ refers to the FC HF energy computed at HB whereas $E(CORR)_{MTA}^{HB}$ is MTA correlation energy computed at HB. The correlation energies in FC at LB and MTA at LB are denoted respectively as $E(CORR)_{FC}^{LB}$ and $E(CORR)_{MTA}^{LB}$. The grafting correction is proposed on the basis of our earlier observation that the difference between MTA and FC energies is almost independent of basis set used.[22] As expected, this correction leads to a better estimation of the total energy and electronic properties. The grafting procedure is now built into the MTA code. The final energy printed out in the MTA code already incorporates the grafting correction. In view of this, in the discussion that follows, the grafted MTA energy will be denoted as $E_{MTA}$.

In the present work, we explore the construction of PES for three test molecules, *viz.* acetylacetone (AA), N-methylacetamide (NMA) and tropolone with the use of FB MTA. The AA, NMA and tropolone molecules contain 15, 12 and 15 atoms respectively. The starting ground- and transition state geometries as well as other few geometries of these molecules are adopted from the articles by Bowman and co-workers.[23-25] On account of the huge computational cost, they have used a modest level of theory and/or basis set. It would be indeed worthwhile to carry out these calculations with high level of theory and large basis sets for a meaningful comparison with the experimental findings. These geometries (ground- and transition state) are further subjected to the geometry optimization at MP2/aug-cc-pVTZ(MP2/aVTZ) level of theory. The number of basis functions for the aVTZ basis set associated with AA, NMA and tropolone are 506, 391 and 592 respectively. The optimized structures of test molecules are depicted in Figure 1 and the respective Cartesian co-ordinates are given in Table TS1 in the Supplementary Information (SI). The single point (SP) FC energy evaluations at MP2 and CCSD(T) level using aVTZ basis set, are carried out for these geometries. Moreover, the harmonic vibrational frequency computations are also done at MP2/aVTZ level of theory in order to confirm their minimal nature. All computations are done by using the Gaussian suite of programs[26] on a 16-core Hewlett-Packard (HP) server-grade machine, unless indicated otherwise.

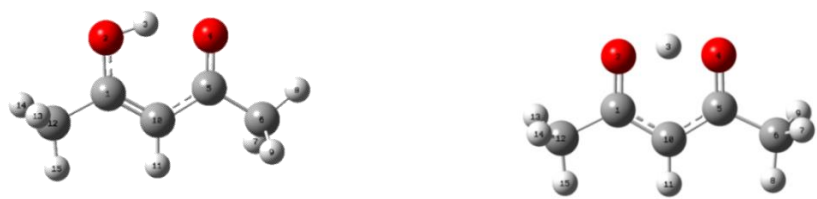

(a) Acetylacetone (AA); Global minimum(GM) (left) and the Transition State (TS) (right)

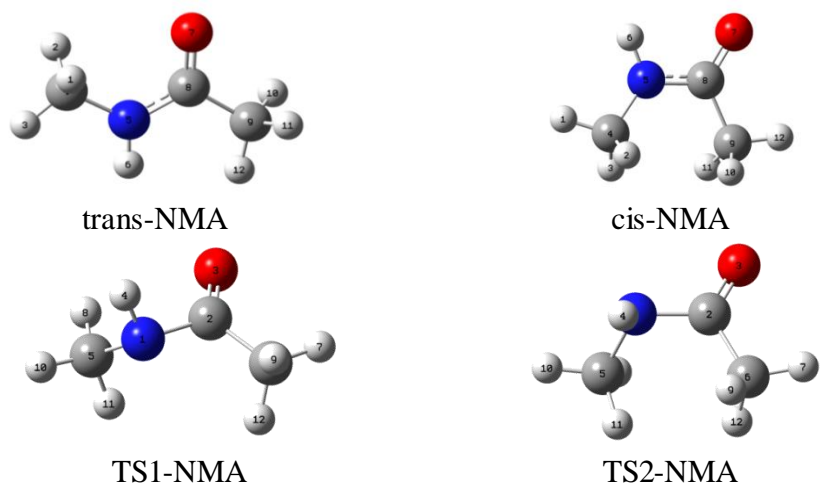

trans-NMA                              cis-NMA

TS1-NMA                                TS2-NMA

(b) *trans* and *cis* structures of N-methylacetamide (NMA) and the two transition states

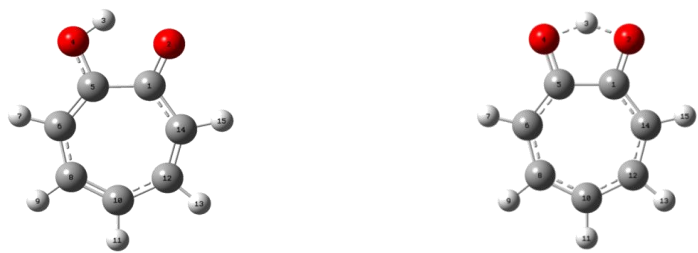

(c) Tropolone: Global Minimum (GM) (left); Transition State (TS) (right)

**Figure 1.** Geometrical structures with the corresponding transition states of (a)Acetylacetone (AA), (b) N-methylacetamide (NMA) and (c) Tropolone molecules optimized at MP2/aVTZ level of theory. See text for details.

The molecules under study, being small in size, are subjected to manual fragmentation into main (M) and overlapping (O) fragments, which are used for MTA computations. The fragmentation schemes for global minimum (GM) geometries of the test molecules are in shown in Figure 2.

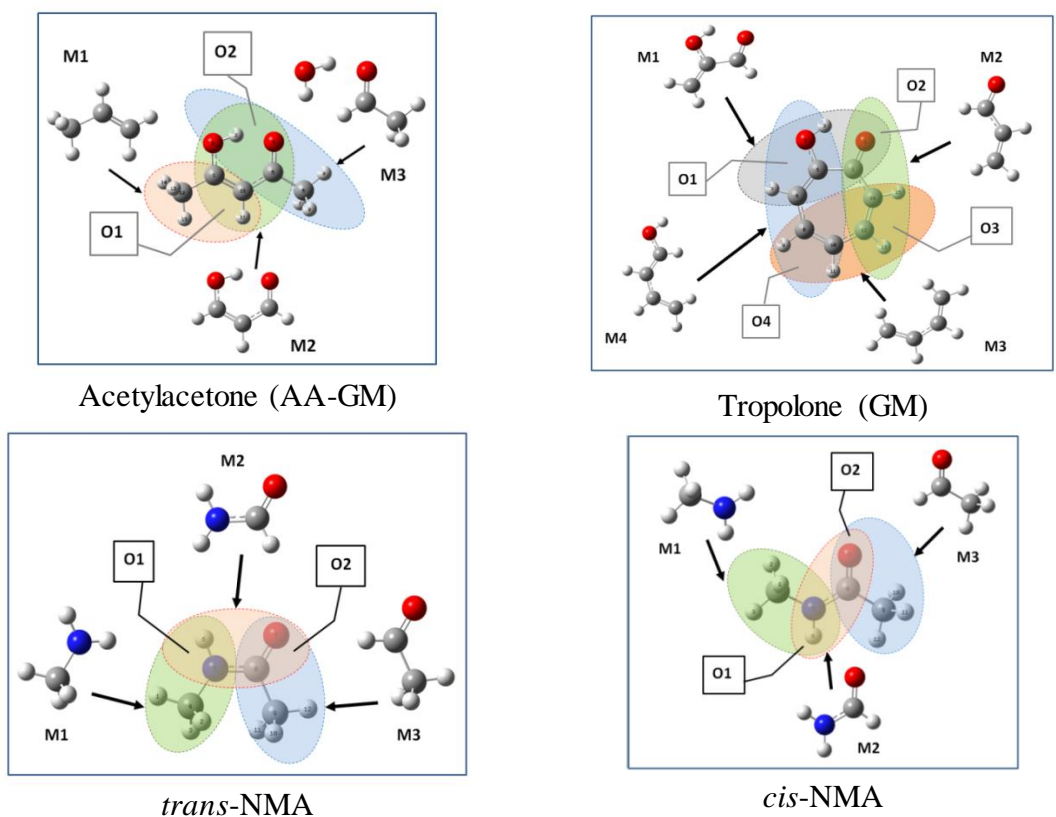

**Figure 2.** Fragmentation schemes implemented in MTA calculations for Acetylacetone (AA-GM), Tropolone (GM), *trans*-NMA and *cis*-NMA. See text for details.

In Figure 2, the main fragments are highlighted in light red, blue and green ovals and are labeled as M1, M2 etc. The respective overlap fragments are labeled as O1, O2 etc. There are maximum three/four main fragments with an average size of half of the molecule and two/three overlapping ones. Identical fragmentation scheme is employed for the TS geometries as well.

The SP MTA energy evaluation for these geometries at MP2/aVTZ and CCSD(T)/aVTZ level are carried out using these fragments. The grafting correction is done employing cc-pVTZ basis set at the respective level of theories. Apart from GM and TS geometries, two other geometries of each of the test molecules are also explored for comparing the MTA energies with their FC counterparts. The Cartesian coordinates of these geometries are specified in the SI.

The SP energy gradient calculations at MP2/aVTZ and CCSD(T)/aVDZ levels are also carried out for one of the geometries (which is not a stationary point on the PES) of each of the three test molecules, with a view to benchmark the accuracy and efficiency of MTA. We report the comparison of the gradients at only one prototype geometry (called as A1, N1, T1 for AA, NMA and Trolopone molecules respectively) for each of the test molecules at FC and MTA levels. A computation for estimating gradients at CCSD(T)/aVTZ level for the geometry A1 of AA is also carried out. Since analytical gradients for CCSD(T) level theory is not available in the Gaussian package,[26] the numerical gradients (as implemented in the Gaussian[26]) are used in the present study.

With this in view, we report herewith the appraisal of the accuracy and efficiency of MTA for reproducing the energies and energy gradients for the test molecules leading a rapid and reliable construction of PES using limited hardware.

The SP energies at MP2 and CCSD(T) for MP2/aVTZ optimized geometries of *cis-* and *trans* conformers, along with two couple of transation states of NMA molecule are displayed in Table 1. Furthermore, the energies for two non-stationary geometries of NMA i.e. N1 and N2 are also reported (*cf.* Table TS2(a) of SI for Cartesian coordinates). Table 1 also reports the signed error, $\Delta = E_{MTA} - E_{FC}$, at the corresponding level of theory, bringing out the excellent agreement of MTA energy with its FC counterpart, the error being uniformly negative and numerically less than 0.7 mH. The wall-clock time shows a significant advantage (a factor of ~3 or so) for MTA *vis-à-vis* the FC for computing the energies at CCSD(T) level. Note that for small test systems no time advantage is seen in the case of MP2 computations. However, a small time advantage is noticed for the largest system, viz. tropolone, with augmented basis sets employing MP2 method.

**Table 1.** Single point Full Calculation ($E_{FC}$) and MTA ($E_{MTA}$) energies[#] (in a.u.) using aVTZ basis for NMA molecule using MP2/aVTZ optimized geometries, along with the (signed) error. The respective wall clock timings (min) are given in parentheses. See text for details.

|  | MP2 | | | CCSD(T) | | |
| --- | --- | --- | --- | --- | --- | --- |
| Geometry | $E_{FC}$ | $E_{MTA}$ | Δ | $E_{FC}$ | $E_{MTA}$ | Δ |
| *Cis* | -248.06688 | -248.06625 | -0.00063 | -248.13930 | -248.13875 | -0.00054 |
| *Trans* | -248.07080 | -248.07024 | -0.00055 | -248.14308 | -248.14263 | -0.00045 |
| TS1-NMA | -248.04284 | -248.04218 | -0.00066 | -248.11685 | -248.11630 | -0.00054 |
| TS2-NMA | -248.03753 | -248.03695 | -0.00058 | -248.11169 | -248.11123 | -0.00046 |
| N1 | -248.06834 | -248.06791 | -0.00043 | -248.14116 | -248.14081 | -0.00035 |
| N2 | -248.05584 | -248.05547 | -0.00037 | -248.12912 | -248.12856 | -0.00055 |
|  | (1.5) | (2.5) |  | (93) | (29) |  |

[#]The MTA energies are calculated employing Eq. (2)

Likewise, for the other two test molecules, *viz.* the *enol* form of AA and tropolone, the MTA- and FC energies at MP2 and CCSD(T) using aVTZ basis set are in shown in Table 2. The Cartesian coordinates of the non-stationary geometries of respective molecules are provided in SI *cf.* Table TS2(a). The energy differences are seen to reduce with the elevation of the level of theory[22] for all the test cases testifying our earlier observation. A time advantage factor between 4 and 5 is noticed for both the molecules at CCSD(T) level theory.

**Table 2:** Single point Full Calculation ($E_{FC}$) and grafted MTA ($E_{MTA}$) energies for the *enol* form of acetylacetone (AA) and tropolone molecules, estimated using aVTZ basis set. The wall clock timings (min) are given in parentheses and the (signed) error is given as Δ. See text for details.

|  | MP2 | | | CCSD(T) | | |
| --- | --- | --- | --- | --- | --- | --- |
| System | $E_{FC}$ | $E_{MTA}$ | Δ | $E_{FC}$ | $E_{MTA}$ | Δ |
| | | *enol* form of acetylacetone (AA) | | | | |
| GM | -345.17691 | -345.17641 | -0.00050 | -345.27104 | -345.27056 | -0.00048 |
| TS | -345.17343 | -345.17299 | -0.00044 | -345.26610 | -345.26567 | -0.00043 |
| A1 | -345.09946 | 345.09910 | -0.00036 | -345.19219 | -345.19181 | -0.00038 |
| A2 | -345.09048 | -345.09023 | -0.00025 | -345.18336 | -345.18290 | -0.00045 |
|  | (6) | (6) |  | (740) | (172) |  |
| | | Tropolone | | | | |
| TS | -420.00049 | -420.00067 | 0.00018 | -420.09604* | -420.09600 | -0.00004 |
| GM | -420.00707 | -420.00667 | -0.00040 | -420.10670* | -420.10642 | -0.00028 |
| T1 | -419.92486 | -419.92624 | 0.00138 | -420.02997 | -420.02948 | -0.00049 |
| T2 | -419.98716 | -419.98826 | 0.00110 | -420.08595 | -420.08663 | 0.00067 |

|  | (21) | (18) | (2040) | (421) |

*Calculation done on 16 cores of a node from NSM Shivay computational facility.

The energy barrier between ground- to transition state is also seen to be well-preserved. For example, in case of AA molecule at CCSD(T)/aVTZ level, the FC barrier of 4.96 mH is well-captured by MTA (4.89 mH). Similarly, for tropolone the FC energy barrier of 10.66 mH is estimated fairly well by the MTA one (10.42 mH).

As mentioned earlier, it is useful to have the energy gradients for different geometries of molecule (which are not the GM or TS) in order to build its PES. However, the FC gradient calculation with high level of theory/basis set is prohibitively difficult. In view of this, an appraisal of MTA for estimating the energy gradients (GR) is carried out.

Since FC gradient calculation is readily possible at the MP2/aVTZ and CCSD(T)/aug-cc-pVDZ (aVDZ) levels, we have used these levels of theory for benchmarking purpose. Table TS2 (b) from the SI compares the MTA gradients at MP2/aVTZ with their FC counterparts. The Cartesian coordinates for one such geometry (N1) of NMA are given in Table TS2(a). Table 3 compares the gradients calculated using FC and MTA at CCSD/aVDZ level theory for the geometry (N1) of NMA. For this purpose, we use a simpler version of grafting correction, viz.

$$GR_{GMTA} = GR_{MTA}^{HB} + (GR_{FC}^{LB} - GR_{MTA}^{LB}) \qquad \ldots(3)$$

The FC and MTA CCSD(T)/aVDZ level gradient results for A1 and T1 are also displayed in Table TS2(b) of SI. All these gradient results show a substantial time advantage factor for MTA gradient calculation, retaining a very good accuracy, especially those gradients with magnitude greater than 0.005 vis-à-vis their FC counterparts.

**Table 3.** Comparison between Full Calculation ($GR_{FC}$) and MTA ($GR_{MTA}$) gradients (in a.u.) calculated at CCSD(T)/aVDZ basis for the geometry (N1) of N-methylacetamide (NMA) molecule. See text for details.

|  | $GR_{FC}$ | | | $GR_{MTA}$ | | |
|---|---|---|---|---|---|---|
| H | -0.0013 | -0.0012 | -0.0030 | -0.0003 | -0.0015 | -0.0027 |
| H | -0.0031 | -0.0021 | 0.0086 | -0.0021 | -0.0024 | 0.0083 |
| H | 0.0016 | 0.0143 | 0.0015 | 0.0029 | 0.0138 | 0.0015 |
| C | 0.0098 | -0.0146 | -0.0085 | 0.0046 | -0.0123 | -0.0084 |
| N | -0.0095 | 0.0222 | 0.0020 | -0.0088 | 0.0213 | 0.0019 |
| H | 0.0018 | -0.0171 | 0.0001 | 0.0020 | -0.0162 | 0.0001 |
| O | -0.0011 | -0.0070 | 0.0013 | -0.0012 | -0.0060 | 0.0013 |
| C | -0.0045 | 0.0090 | -0.0029 | -0.0033 | 0.0096 | -0.0029 |
| C | -0.0061 | -0.0183 | -0.0004 | -0.0066 | -0.0193 | -0.0003 |
| H | 0.0061 | 0.0041 | -0.0072 | 0.0061 | 0.0043 | -0.0071 |
| H | 0.0051 | 0.0000 | 0.0055 | 0.0051 | 0.0002 | 0.0055 |
| H | 0.0012 | 0.0107 | 0.0029 | 0.0013 | 0.0107 | 0.0029 |
| | **0.0222*** | **0.0081#** | **355$** | **0.0213*** | **0.0078#** | **95$** |

*Maximum gradient; #RMS gradient. $Wall clock time (min)

Moving ahead, the MTA gradient calculation at CCSD(T)/aVTZ level of theory for the smallest test case i.e. NMA is reported in Table 4. The elapsed time for this computation is somewhat large, viz. 1636min. Although benchmarking with the respective FC results is not feasible, we trust that the results are accurate with significant time advantage employing off-the-shelf hardware. The CCSD(T)/VTZ level of theory is used for grafting correction. As this is a prototype computation, a single 16 core machine was employed for this calculation. However, for production jobs, the multiple nodes can be easily harnessed being MTA code is highly parallel.

**Table 4.** The MTA-Gradients ($GR_{MTA}$) in a.u. calculated at CCSD(T)/aVTZ level of theory for the geometry (N1) of the NMA molecule. See text for details.

| | $GR_{MTA}$ | | |
|---|---|---|---|
| H | 0.0033 | 0.0006 | 0.0037 |
| H | 0.0019 | 0.0000 | 0.0017 |
| H | 0.0014 | 0.0055 | 0.0016 |
| C | 0.0060 | -0.0115 | -0.0083 |
| N | -0.0056 | 0.0235 | 0.0019 |
| H | 0.0017 | -0.0098 | 0.0003 |
| O | -0.0013 | -0.0239 | 0.0014 |
| C | -0.0076 | 0.0168 | -0.0030 |
| C | -0.0048 | -0.0182 | -0.0007 |
| H | 0.0016 | 0.0019 | -0.0001 |
| H | 0.0009 | -0.0020 | -0.0015 |
| H | 0.0020 | 0.0180 | 0.0032 |
| | **0.0239*** | **0.0085#** | **1636$** |

*Maximum gradient; #RMS gradient. $Wall clock time (min)

Table 4 brings out the possibility of applying the MTA for obtaining the energy gradients at such high level of theory viz. CCSD(T) using aVTZ basis set on an off-the-shelf hardware.

The systems under consideration, though apparently rather small, are computationally heavy at the CCSD(T)/aVTZ level. Thus, FC gradient calculations for these molecules may be possible only by investment of very large hardware and wall clock time. However, for spatially extended large molecule, MTA can be only the possible solution.

In summary, the present work has reported an application of MTA for the construction of PES at MP2- and CCSD(T)/aVTZ level of theory. The energies of ground- and transition states, as well as other some other geometries are probed using the MTA methodology. For all the test molecules, the energetics are accurately estimated, with the error vis-à-vis the respective full calculation being smaller than 1 millihartree. The energy barriers from the ground- to transition state are also projected accurately for all the test cases.

Furthermore, the accuracy of the MTA for computing the energy gradients is critically accessed. All the energy gradients at CCSD(T)/aug-cc-pVDZ level are assessed correctly within limited time using single 16- core machine. With the inherently parallel nature of MTA, we expect a significant time saving with the use of multiple machines in parallel. More time advantage can also be reaped using other sophisticated parallel *ab initio* packages like ORCA[27] and PSI4[28]. The customization of MTA to accommodate these software is underway. We trust that the present work paves the way toward an efficient and economic construction of PES for large molecules using high level *ab initio* methods.

**Acknowledgments**

SRG acknowledges the support from the National Supercomputing Mission (NSM) under the Project [CORP: DG: 3187] and computational resources made available on the ParamShivay facility at IIT-BHU, Varanasi. AN is thankful to Professor Joel Bowman for fruitful discussions.